\documentclass[10pt,a4paper,twocolumn]{article}
\usepackage[utf8]{inputenc}
\usepackage{amssymb}
\usepackage{siunitx}
\usepackage{comment}
\usepackage{makecell}
\usepackage{bm}
\usepackage{placeins}
\usepackage{natbib}
\usepackage{graphicx}
\usepackage{hyperref}
\usepackage{geometry}
\usepackage{xcolor}
\usepackage[normalem]{ulem}

\def\aj{{AJ}}                   
             
\def\apj{{ApJ}}                 
                
\def\apjs{{ApJS}}               
           
\def\apss{{Ap\&SS}}             
\def\aap{{A\&A}}                
\def\aapr{{A\&A~Rev.}}          
\def\aaps{{A\&AS}}

\def\memras{\ref@jnl{MmRAS}}            
\def\mnras{{MNRAS}} 
\def\ssr{{Space~Sci.~Rev.}}

\def\jgr{{J.~Geophys.~Res.}}

\DeclareSIUnit\jansky{Jy}
\DeclareSIUnit\parsec{pc}
\DeclareSIUnit\beam{beam}

\def\degr{\hbox{$^\circ$}}
\def\arcmin{\hbox{$^\prime$}}
\def\arcsec{\hbox{$^{\prime\prime}$}}

\title{Characterization of the decametre sky at sub-arcminute resolution}
\author{C. Groeneveld,
        R.~J. van Weeren,
        E. Osinga,
        W.~L. Williams,
        J.~R. Callingham,\\
        F. de Gasperin,
        A. Botteon
        T. Shimwell,
        F. Sweijen,
        J.~M.~G.~H.~J. de Jong,\\
        L.~F. Jansen,
        G.~K. Miley,
        G. Brunetti,
        M. Br\"{u}ggen,
        H.~J.~A. R\"{o}ttgering}
\date{\today}

\begin{document}

\twocolumn
\maketitle

{\bf The largely unexplored decameter radio band (10-30~MHz) provides a unique window for studying a range of astronomical topics, such as auroral emission from exoplanets, inefficient cosmic ray acceleration mechanisms, fossil radio plasma, and free-free absorption. 
The scarcity of low-frequency studies is mainly due to the severe perturbing effects of the ionosphere. %
Here we present a calibration strategy that can correct for the ionosphere in the decameter band. We apply this to an observation from the Low Frequency Array (LOFAR) between 16--30\,MHz. The resulting image covers 330 square degrees of sky at a resolution of 45\arcsec{}, reaching a sensitivity of 12\,mJy~beam$^{-1}$. Residual ionospheric effects cause additional blurring ranging between 60 to 100\arcsec{}. This represents an order of magnitude improvement in terms of sensitivity and resolution compared to previous decameter band observations. %
{In the region we surveyed, we have identified four fossil plasma sources. These rare sources are believed to contain old, possibly re-energised, radio plasma originating from previous outbursts of active galactic nuclei.  At least three of them are situated near the center of low-mass galaxy clusters. Notably, two of these sources display the steepest radio spectral index among all the sources detected at 23\,MHz. This indicates that fossil plasma sources constitute the primary population of steep-spectrum sources at these frequencies, emphasising the large discovery potential of ground-based decameter observations.}

}

The first detection of extraterrestrial radio waves was made in the decameter band by Karl G. Jansky \citep{jansky1932directional}. After these initial observations, most studies of the radio sky were carried out at higher frequencies due to greater observational and technical advantages, such as increased resolution, lower sky temperature, and improved ability to be calibrated. Therefore, the decameter band remains a relatively unexplored spectral window. There has recently been a resurgence of interest in the decameter band because it is a unique window for studying several  astrophysical sources and radiation mechanisms that are inaccessible at higher frequencies. Some classes of radio sources are known to emit primarily at low frequencies with a sharp spectral cutoff. Others have ultra-steep radio spectra throughout the radio frequencies at which they have been observed. Furthermore, several absorption processes which suppress low-frequency emission become more dominant in the decameter band \citep[e.g., free-free absorption, synchrotron self-absorption; ][]{1986ApJ...301..813O,callingham2015}.
Decameter observations can provide critical constraints on the underlying emission and absorption processes.

\begin{figure*}[t]
    \centering
    \includegraphics[width=0.85\textwidth]{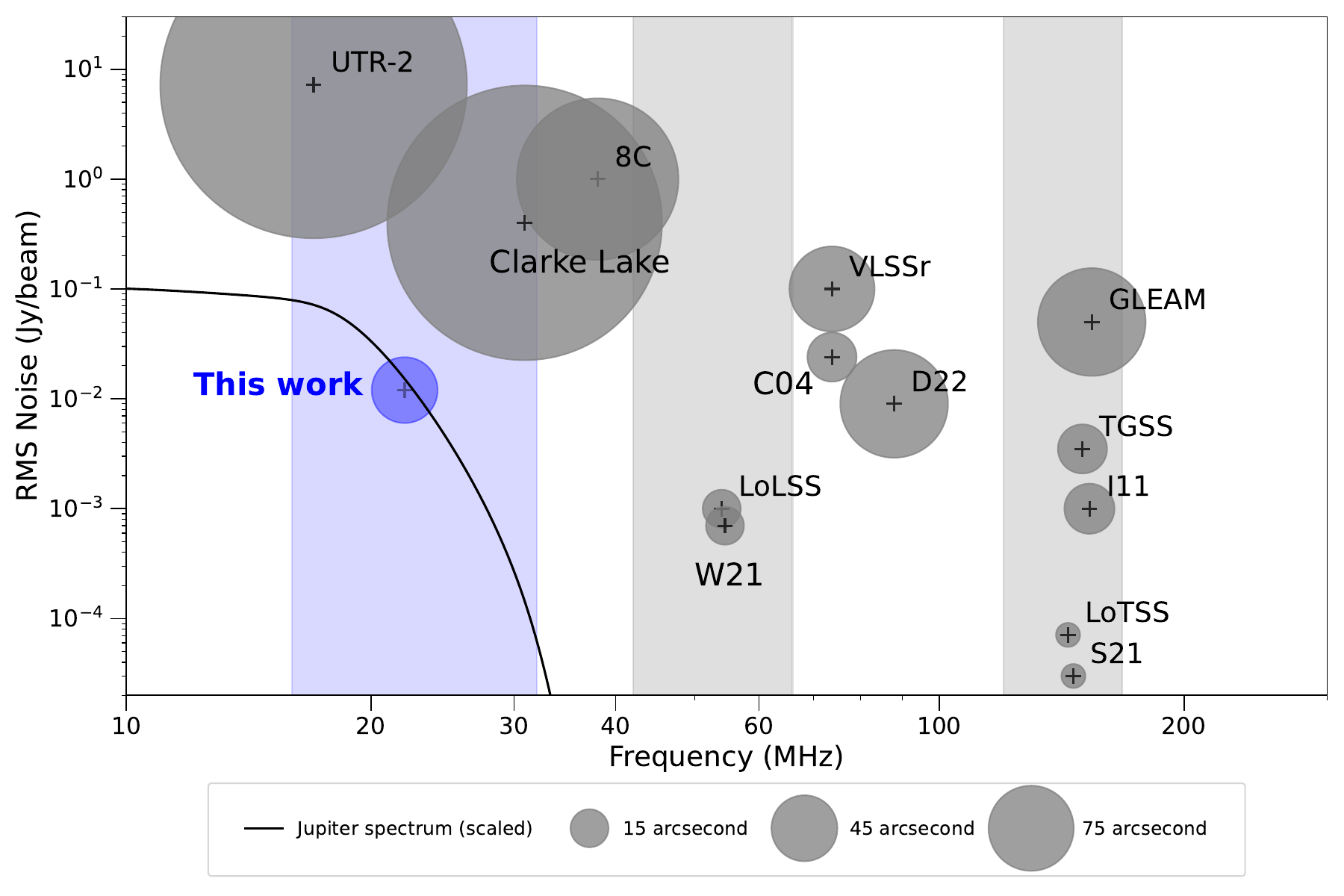}
    \caption{Comparison of the RMS noise and central frequency of previous low-frequency observations. Highlighted in blue is the wavelength range covered by the data described in this work. A Jupiter spectrum (extracted from \cite{zarka2007}) scaled to match the sensitivity of this work, is included to illustrate the importance of the decameter band for detecting electron-cyclotron maser emission. The size of the circles represents the resolution of the survey. For the LOFAR surveys (LoTSS, LoLSS), the frequency ranges are indicated with grey-shaded bands. The included surveys are detailed by \citet{UTRmain}, \citet{clarkelake}, \citet{eightsee}, \citet{lolls}, \citet{williams21}, \citet{cohen2004}, \citet{duchesne}, \citet{gleam}, \citet{tgss}, \citet{intema2011}, \citet{lotss_I}, \citet{Lotss-deepfield}}
    \label{fig:surveycompare}
\end{figure*}

Coherent radio emission from stellar systems can be bright in the decameter band \citep{callingham2021}. This coherent emission is produced by plasma or electron cyclotron maser instability (ECMI) and is often generated in the coronae of stars. ECMI emission can also result from the interaction of a stellar or planetary companion with the magnetosphere of the primary body \citep[e.g.][]{callingham2021}. These auroral emissions are ubiquitous for planets in our Solar System which harbor strong magnetic fields, including the Earth. In the case of Jupiter, one prominent pathway by which ECMI emission can be induced is the interaction of Io with Jupiter's magnetosphere \citep[][]{jupiterio_aurora}. The radio spectrum from ECMI emission has a strong cutoff in the spectrum. %
For example, %
Jupiter is four orders of magnitude brighter in the decameter wavelength band than at frequencies above 30~MHz \citep{zarka2004}, which makes decameter observations uniquely capable for observing similar systems \citep[e.g. in star--planet interaction systems, see][]{harish}. Pushing the low-frequency limit is important to enable the detection of auroral emission from exoplanets with weaker magnetic fields since the cutoff frequency is directly proportional to the magnetic field strength of the emitting body. %
{The decameter radio band is also sensitive to the lower energy synchrotron-emitting cosmic ray electrons that trace long timescales and inefficient acceleration mechanisms \citep[such as turbulent re-acceleration][]{2005MNRAS.357.1313C}. For example, the lobes of radio galaxies will steepen with age when the active galactic nucleus (AGN) ceases its activity, due to synchrotron and Inverse Compton losses. When this old ($\sim 10^{7-9}$\,yr) AGN radio plasma is located in the hot plasma that permeates galaxy clusters, it can be re-energised by adiabatic compression or other re-acceleration mechanisms \citep{fossil_plasma,2017SciA....3E1634D,brienza_natast}. This produces sources with complex morphologies that almost exclusively emit at low frequencies \citep{2002MNRAS.331.1011E}. Decameter observations thus provide a unique view of the complex life cycle of cosmic rays in galaxy clusters.}

\begin{figure*}
    \centering
    \vspace{-3cm}
    \includegraphics[width=\textwidth]{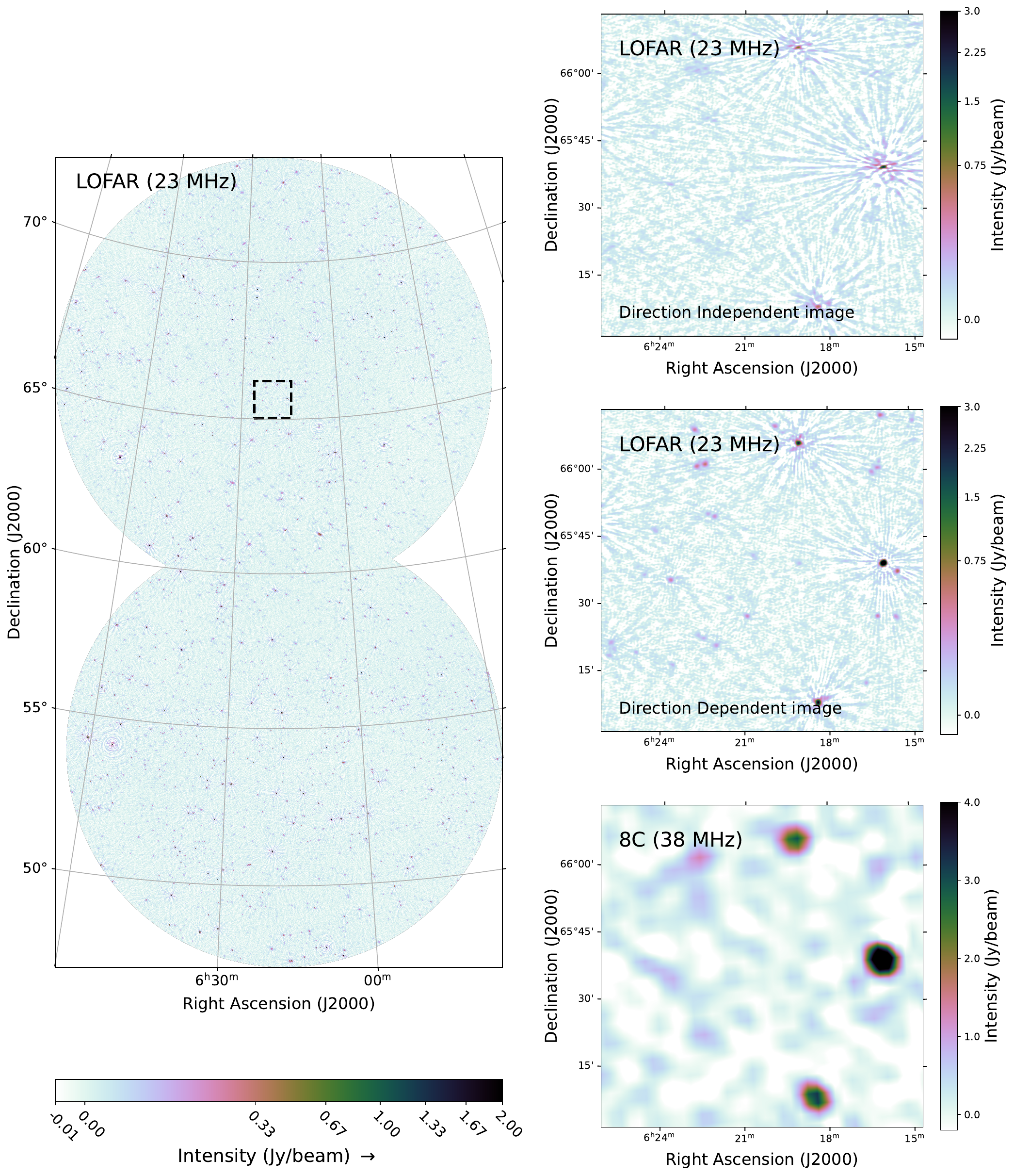}
    \caption{\textit{Left:} Full image of our observed region at a central frequency of 23\,MHz. The image is clipped at a field of view of 7 degrees, which corresponds with the FWHM of the LOFAR beam at these frequencies. The black dashed rectangle corresponds to the region highlighted in the three panels on the right. \textit{Right:} Comparison between direction-independent calibration (top) and the facet-based direction-dependent calibration (middle) of our data. The prominent 'spikes' in the direction-independent image are caused by imperfect corrections of the ionosphere, as the ionosphere changes rapidly across the field of view. With direction-dependent calibration, the full field is divided into smaller facets that are individually corrected for the ionosphere, thus mitigating most of these spikes. As a comparison, the 8C image of the same area is given (bottom).}
    \label{fig:fullfov}
\end{figure*}

While there are several important astronomical phenomena that can only be probed in the decameter band, there are three problems that must be overcome when observing at such frequencies. First, because of the long wavelengths involved, the resolution that can be obtained is inherently limited. Achieving subarcminute resolution requires baselines of the order of $100$\,km. Second, the decameter band often suffers from strong radio frequency interference \citep[RFI; e.g.,][]{lofar_RFI,2017MNRAS.467.2274S}, particularly from shortwave radio communications. Third, the ionosphere significantly influences the propagation of radio waves, which severely complicates the necessary high-resolution imaging. Variations in the ionosphere's total free electron content (TEC) cause variable phase delays in the arriving wavefront, in turn blurring radio images. In addition, the interactions of these electrons with magnetic fields in the ionosphere cause differential Faraday rotation. Both effects are also time (on timescales of $\sim$\,seconds) and direction (on scales of a degree on the sky) dependent. The phase delays from TEC differences scale linearly with the wavelength of incident radio waves, and Faraday rotation scales with the wavelength squared \citep[e.g.,][]{ULFionosphere}. Therefore, when observing at longer wavelengths, these effects become considerably more severe, ultimately becoming uncorrectable when approaching the plasma frequency of the ionosphere at around 3--10\,MHz \citep[e.g.][]{1966ApJ...143..227E}. Given the above challenges, space-based radio arrays at frequencies below 30\,MHz have been considered \citep[e.g.,][]{olfar_roadmap}. 
Only a few decameter ground-based studies have been carried out that were sensitive enough to detect more than several dozen sources \citep[e.g.,][]{1988ApJS...68..715K,UTRmain}. Surveys covering a significant part of the sky have been performed by the Dominion Radio Astrophysical Observatory \citep[22\,MHz;][]{drao_22mhz} and the Ukrainian T-Shaped Radio Telescope Second Modification \citep[UTR-2, 10-25\,MHz;][]{UTRmain}. The resolutions of these surveys range from approximately half to a few degrees, with noise levels of a few Jansky per beam or higher. Slightly outside the decameter band, the Eighth Cambridge Catalogue of Radio Sources \citep[8C;][]{eightsee} at 38\,MHz cataloged the sky above a declination of +60\degr{} at a resolution of approximately 4.5\arcmin. This survey achieved a noise level of 0.3\,Jy\,beam$^{-1}$. 

Recently, significant progress in the calibration of low-frequency radio observations has been made with the Low-Frequency Array \citep[LOFAR;][]{lofarpaper}, a pan-European radio array sensitive to frequencies between 10 and 240\,MHz. %
The core of LOFAR comprises baselines up to 120\,km in the Netherlands. Previous studies with LOFAR consist of, among others, the LOFAR Two-meter Sky Survey \citep[LoTSS,120-168\,MHz][]{lotss_I,2022A&A...659A...1S} and the LOFAR LBA Sky Survey \citep[LoLSS, 42-66\,MHz,][]{lolls,lollsDR1}. 
The sensitivity, resolution, and frequency coverage of these observations, along with previous low-frequency work, are shown in \autoref{fig:surveycompare}.

In this work, we demonstrate the feasibility of sensitive subarcminute resolution decameter band observations using a calibration strategy to correct for the severe ionospheric image blurring that occurs below 30~MHz.
We apply this to a 5\,hr dual-beam LOFAR observation between 16 and 30\,MHz, going three times lower in frequency than LoLSS. This calibration strategy allows us to create a sub-arcminute resolution image of the decameter sky for the first time (see \autoref{fig:fullfov}).
To verify that the ionospheric conditions during our observations were typical, we analyzed  observations of the primary calibrator over several nights. These demonstrated that the ionospheric conditions present during our observations were typical, based on the S4 index (see Methods). 

\begin{figure}
    \centering
    \includegraphics[width=\linewidth]{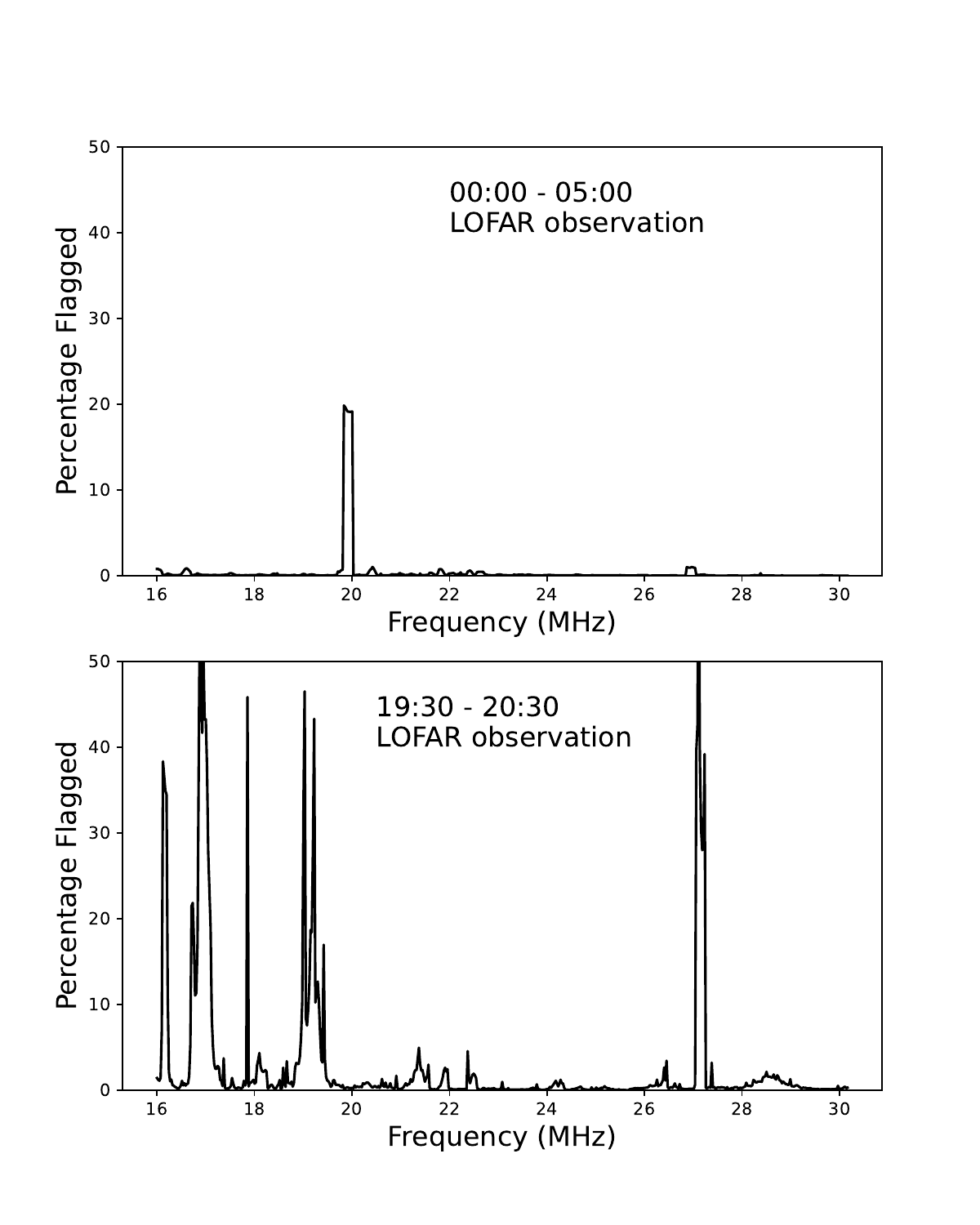}
    \caption{Fraction of LOFAR data flagged due to RFI as a function of frequency. The top image shows the observation used in this paper and the bottom image shows an observation taken in the early evening local time. There was significantly more RFI during the latter observation, by about a factor of 7. %
    Similar bad RFI conditions are experienced in the morning and during the day.}
    \label{fig:flagged}
\end{figure}

Our observing strategy consisted of simultaneously observing a bright primary calibrator (3C\,196) and the target fields. By scheduling the observation after midnight, we minimized RFI caused by the internal reflection of terrestrial RFI by the ionosphere, which is significantly worse during the day, as ionizing radiation from the Sun increases the column density of ions in the ionosphere. The remaining RFI is excised at high time (1\,second) and frequency (3.05~\,kHz) resolution. The overall amount of data lost due to RFI was less than 10\% (see \autoref{fig:flagged}).

By adapting the method of \cite{facet_calibration,lolls}, using individual facets calibrated for direction-dependent effects, we show that this approach can be even successful in the decameter radio band.  First, instrumental effects are corrected by using 3C\,196 as a calibrator source. Next, differential Faraday rotation and phase corrections for first and third-order ionospheric effects were obtained for the whole field of view, after which the full field was imaged. %
This preliminary image reveals about 30 bright sources, although the majority of sources are severely blurred and smeared out, due to direction-dependent effects of the ionosphere.
Following this step, we split up our field of view into several smaller ``facets'' and self-calibrated each facet individually, against a starting model obtained from the 151~MHz TGSS ADR1 \citep{tgss} survey. This yields an improved image and model of the sky, partly corrected for direction-dependent effects. We repeated this step three more times, increasing the number of facets and iteratively constructing a better model of the sky, which in turn allowed us to calibrate on facets containing less flux.
Based on the activity of the ionosphere, 25 facets were calibrated and used in order to create a sharp image of the full field of view.

The resulting image has a resolution of 45\arcsec{} at a central frequency of 23\,MHz, and reaches a central noise level of 12\,mJy\,beam$^{-1}$, which is more than three orders of magnitude in improvement compared to previous work below 30~MHz \citep[e.g. ][]{UTRmain,drao_22mhz}. The full region mapped is presented in \autoref{fig:fullfov}. The image covers an area of 330 square degrees, consisting of the sky area within 7 degrees of the pointing center of each observation. This corresponds to 0.6 times the full width at half maximum (FWHM) of the primary beam of the used array at 23\,MHz. Beyond this radius, the image quality rapidly decreases due to the sparsity of the direction-dependent corrections. 
The improvement due to our facet-based, direction-dependent approach can be seen in the cutout shown in \autoref{fig:fullfov}. While the image quality is greatly improved, some  artifacts remain around the brighter sources due to imperfections in the ionospheric corrections.
These imperfections can be modeled with a Gaussian blurring of $\SI{60}{\arcsec} $ for brighter sources, up to $\SI{100}{\arcsec}$ for fainter sources.

A source catalog was constructed from the image using pyBDSF \citep{pybdsf}, with flux densities measured in apertures having radii of 2\,arcmin. This ensures that the measured flux densities are accurate regardless of the shape of the point spread function. The source catalog can be found online, or attached to this paper. We detected a total of 2863 sources within the 330 square degrees coverage of the image and a signal-to-noise ratio of at least 7, defined by the ratio between the flux density of a source divided by the local RMS. 
This corresponds to a source density of 8.7 sources per square degree and an improvement of two orders of magnitude compared with the UTR-2 and DRAO 22~MHz surveys, the benchmark surveys at these frequencies. %
Our measured source positions are consistent to within $< \SI{10}{arcsec}$ with those from TGSS and there appears to be no systematic flux-density scale offset compared with previous source catalogs (see Methods). 

\begin{figure*}
    \centering
    \includegraphics[width=0.85\linewidth]{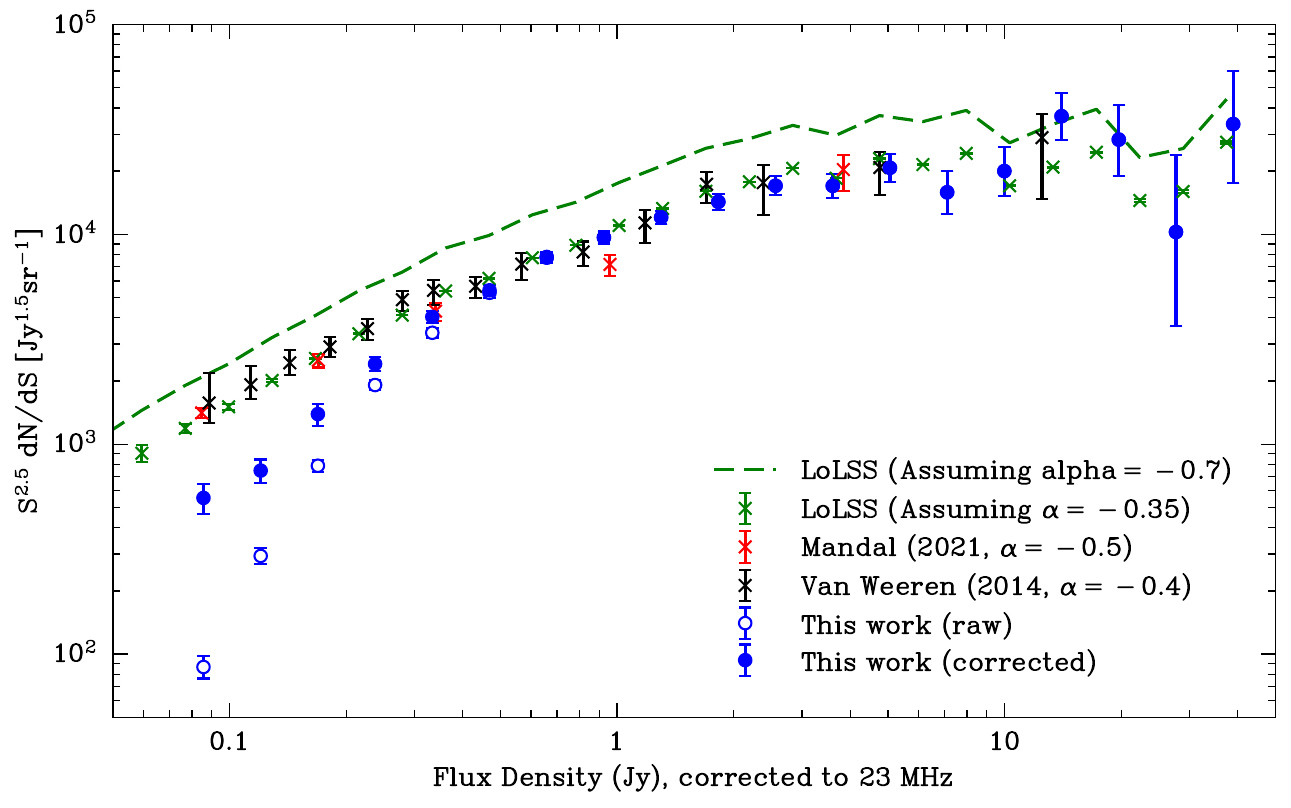}
    \caption{Euclidean normalized source counts in the surveyed area. Included are the source counts as constructed in this work, from \cite{vanWeerensourcecounts}, from \cite{2021A&A...648A...5M}, and from \cite{lolls}, scaled with both a $\alpha=-0.7$ and $\alpha=-0.35$ power law. The corrected source counts are reported in \autoref{tab:ensc}}
    \label{fig:ensc}
\end{figure*}

\begin{table*}[hb]
    \centering
    \begin{tabular}{lccr}
        \textbf{S[Jy]} & \textbf{\textless} $\bm A_{vis}$ \textbf{\textgreater} & \textbf{dN} & \makecell{$\text{\bf S}^{\bm{2.5}} \text{\bf dN/dS}$ \\ $\text{ \bf [Jy}^{\bm{1.5}}\text{\bf sr}^{\bm{-1}}${\bf ]}} \\ \hline 
        $ 0.10$--$ 0.14$ & $ 0.78$ & $   85^{ 10}_{  9}$ & $575.41^{79.37}_{73.17}$ \\
        $ 0.14$--$ 0.19$ & $ 0.82$ & $  151^{ 13}_{ 12}$ & $766.83^{108.39}_{105.16}$ \\
        $ 0.19$--$ 0.27$ & $ 0.86$ & $  254^{ 17}_{ 16}$ & $1421.55^{139.88}_{136.02}$ \\
        $ 0.27$--$ 0.37$ & $ 0.91$ & $  355^{ 20}_{ 19}$ & $2389.14^{193.47}_{188.75}$ \\
        $ 0.37$--$ 0.51$ & $ 0.95$ & $  392^{ 21}_{ 20}$ & $4067.12^{243.19}_{233.77}$ \\
        $ 0.51$--$ 0.71$ & $ 0.98$ & $  365^{ 20}_{ 19}$ & $5335.53^{300.06}_{285.33}$ \\
        $ 0.71$--$ 0.99$ & $ 0.99$ & $  339^{ 19}_{ 18}$ & $7943.28^{456.65}_{432.58}$ \\
        $ 0.99$--$ 1.37$ & $ 1.00$ & $  254^{ 17}_{ 16}$ & $9669.55^{645.71}_{606.30}$ \\
        $ 1.37$--$ 1.90$ & $ 1.00$ & $  194^{ 15}_{ 14}$ & $12043.30^{928.42}_{863.85}$ \\
        $ 1.90$--$ 2.63$ & $ 1.00$ & $  140^{ 13}_{ 12}$ & $14181.75^{1303.08}_{1196.99}$ \\
        $ 2.63$--$ 3.65$ & $ 1.00$ & $  103^{ 11}_{ 10}$ & $17036.31^{1850.14}_{1675.62}$ \\
        $ 3.65$--$ 5.07$ & $ 1.00$ & $   64^{  9}_{  8}$ & $17287.51^{2443.68}_{2154.69}$ \\
        $ 5.07$--$ 7.03$ & $ 1.00$ & $   47^{  8}_{  7}$ & $20735.03^{3489.72}_{3012.61}$ \\
        $ 7.03$--$ 9.74$ & $ 1.00$ & $   22^{  6}_{  5}$ & $15852.50^{4157.46}_{3351.40}$ \\
        $ 9.74$--$13.51$ & $ 1.00$ & $   18^{  5}_{  4}$ & $21184.62^{6273.15}_{4942.09}$ \\
        $13.51$--$18.74$ & $ 1.00$ & $   18^{  5}_{  4}$ & $34601.74^{10246.20}_{8072.12}$ \\
        $18.74$--$25.99$ & $ 1.00$ & $    9^{  4}_{  3}$ & $28258.10^{12943.78}_{9227.04}$ \\
        $25.99$--$36.05$ & $ 1.00$ & $    2^{  3}_{  1}$ & $10256.69^{13632.74}_{6605.18}$ \\
        $36.05$--$50.00$ & $ 1.00$ & $    4^{  3}_{  2}$ & $33505.31^{26632.11}_{15990.71}$ \\
    \end{tabular}
    \caption{Table containing the Euclidean normalised source counts, corresponding with Figure 4 in the text. Source counts are represented by the expected count $\pm$ standard error.}
    \label{tab:ensc}
\end{table*}

The source catalog provided in this paper is used for constructing Euclidean normalized source counts \citep{dezotti2010} in \autoref{fig:ensc}.
Our source counts show similar behavior to source counts obtained at higher frequencies, with a gradual decrease in the counts towards lower integrated flux densities.
From the figure, it is clear that the catalog is reasonably complete down to a flux density of around 200 mJy, %
below which the correction factors for incompleteness become significant.
With an assumed blurring of around a factor 3 between the peak flux density and integrated flux density (see methods), this corresponds to a peak flux density of around 70 mJy, which indeed corresponds to slightly less than $7\sigma$ of our reported noise.
The main result from these source counts is that we find evidence for flattening of the average spectral index of the source population at 23~MHz. The spectral index required to match the 54~MHz source counts from LoLSS \citep{lollsDR1} (-0.35$\pm 0.1$) is flatter than the spectral index required to match the 1.4~GHz source counts from \cite{dezotti2010} to LoLSS \citep[-0.6$\pm 0.1$, Fig. 21 in][]{lollsDR1}. For reference, \autoref{fig:ensc} also contains the source counts from \cite{lolls}, but scaled to the canonical spectral index of -0.7 instead \citep[e.g.][]{condon_spix}, illustrating that the source population reflects a flatter spectral index. 
This is in agreement with the findings in \cite{vanWeerensourcecounts}, where similar flattening at low frequencies has been found.
Furthermore, this is also consistent with our finding that the median spectral index of our $>$ 200 mJy integrated flux sources is $-0.50\pm 0.01$ measured between 144~MHz and 23~MHz, which is also steeper than the canonical value of -0.7 between 144~MHz and 1.4~GHz.
This curvature is also visible in Fig. S4, where the majority of the nine brightest sources are shown to host curved spectra between 1.4~GHz and 23~MHz, and none of these nine sources show inverted spectra. The curvature can be caused via absorption effects, which flattens the radio spectra at lower frequencies \citep{1986ApJ...301..813O,callingham2015}, or via spectral aging, which causes steepening at higher frequencies \citep{harwood15}.

Given that 98\% of our detected sources are too faint to be detected by other decametric surveys, we compare our results to the 38\,MHz 8C survey, see \autoref{fig:fullfov}. Our results  yield a significant improvement in both resolution and sensitivity over this survey. About 76\% of the sources cataloged by our LOFAR decameter observations were undetected by the 8C survey, but are revealed in the LOFAR decameter observations due to the significantly higher sensitivity. In addition, several cataloged individual 8C sources are shown to be multiple sources, which are blended in 8C due to its worse resolution (4 arcmin). Of the 334 sources from 8C that are within the sky coverage of our observations, 323 are detected by LOFAR and 11 are absent. Eight of the 11 undetected 8C sources are probably spuriously detected by 8C, given their low signal-to-noise ratio in the 8C catalog. The other three sources are visible in the final image but were rejected in the catalog since these sources have a signal-to-noise ratio below seven.

\begin{figure*}
\vspace{-2.5cm}
    \centering
    \includegraphics[width=0.9\textwidth]{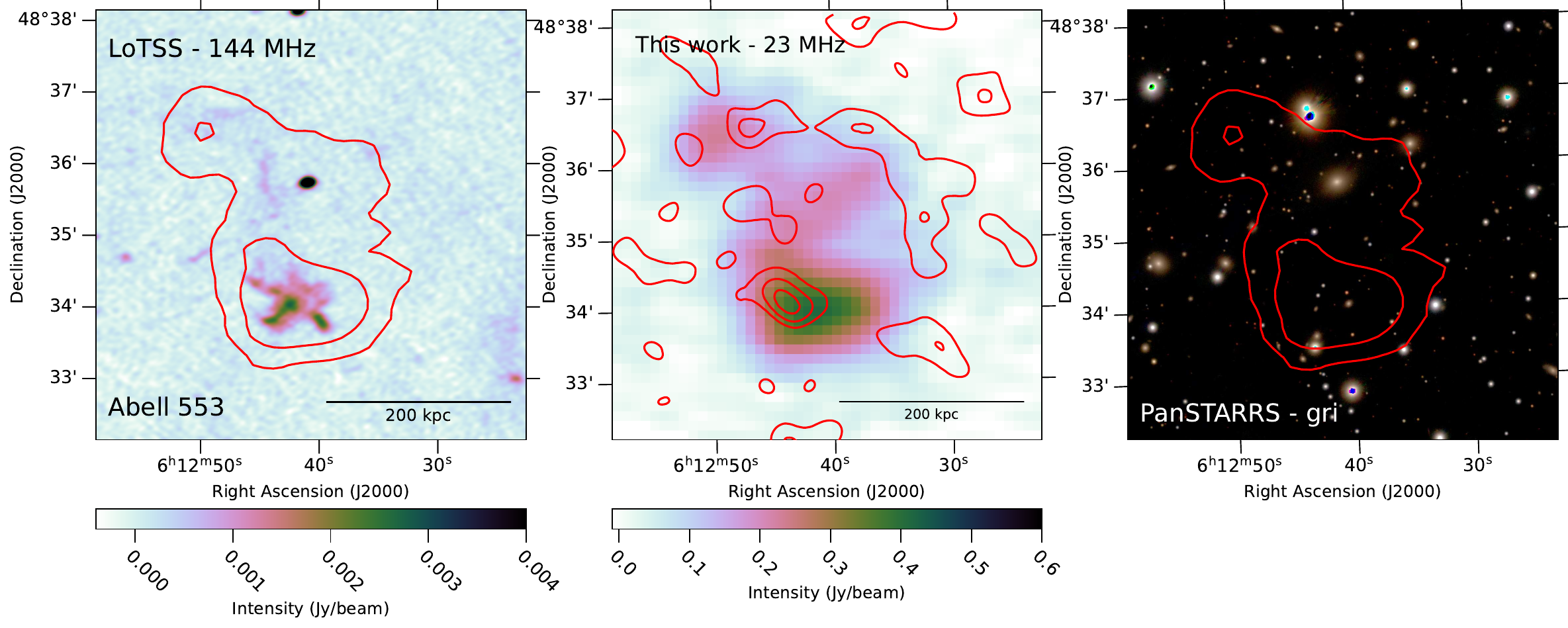}\\
    \includegraphics[width=0.9\textwidth]{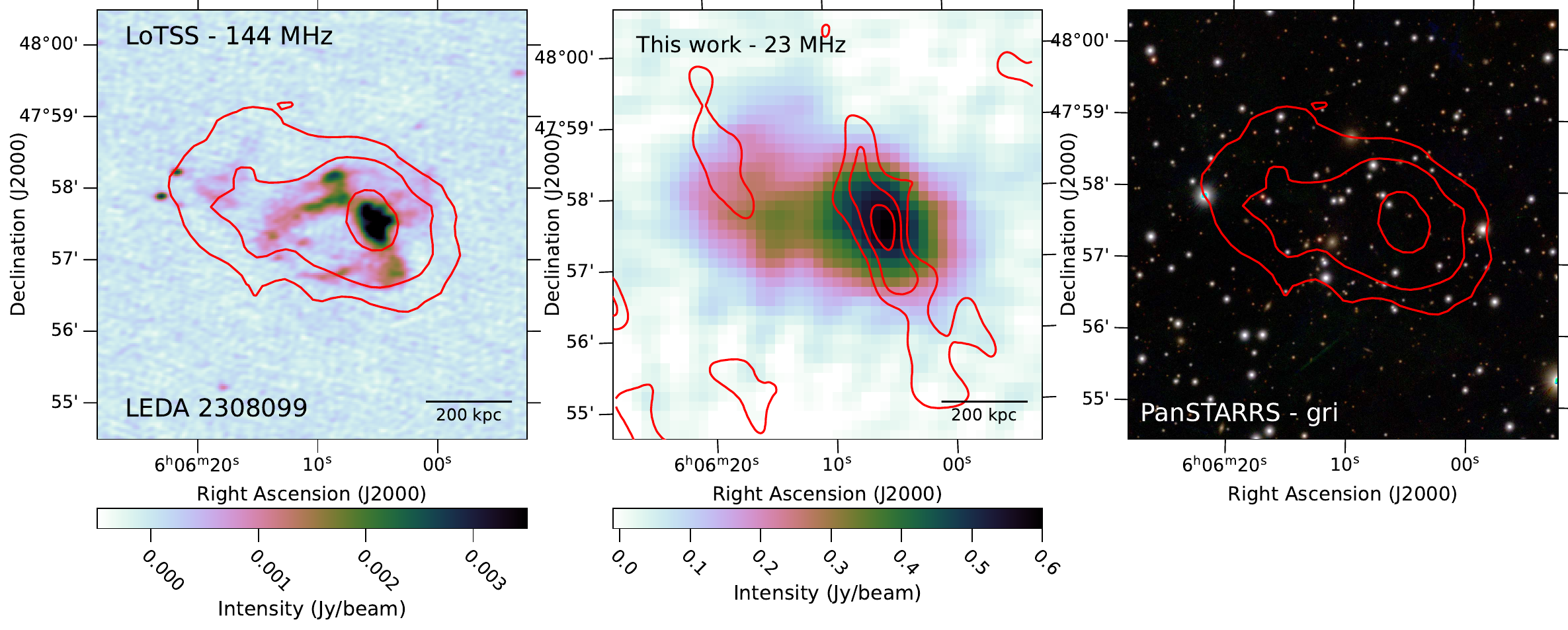}\\
    \includegraphics[width=0.9\textwidth]{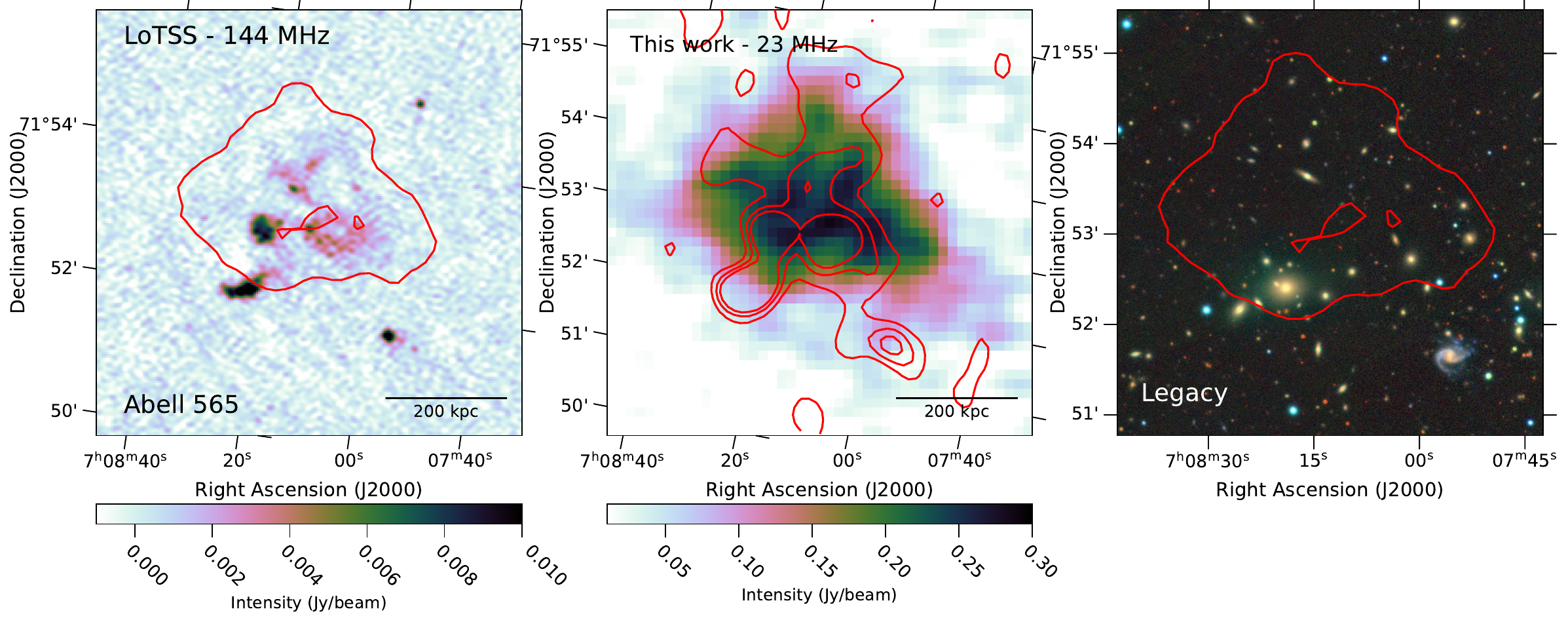}\\
    \includegraphics[width=0.9\textwidth]{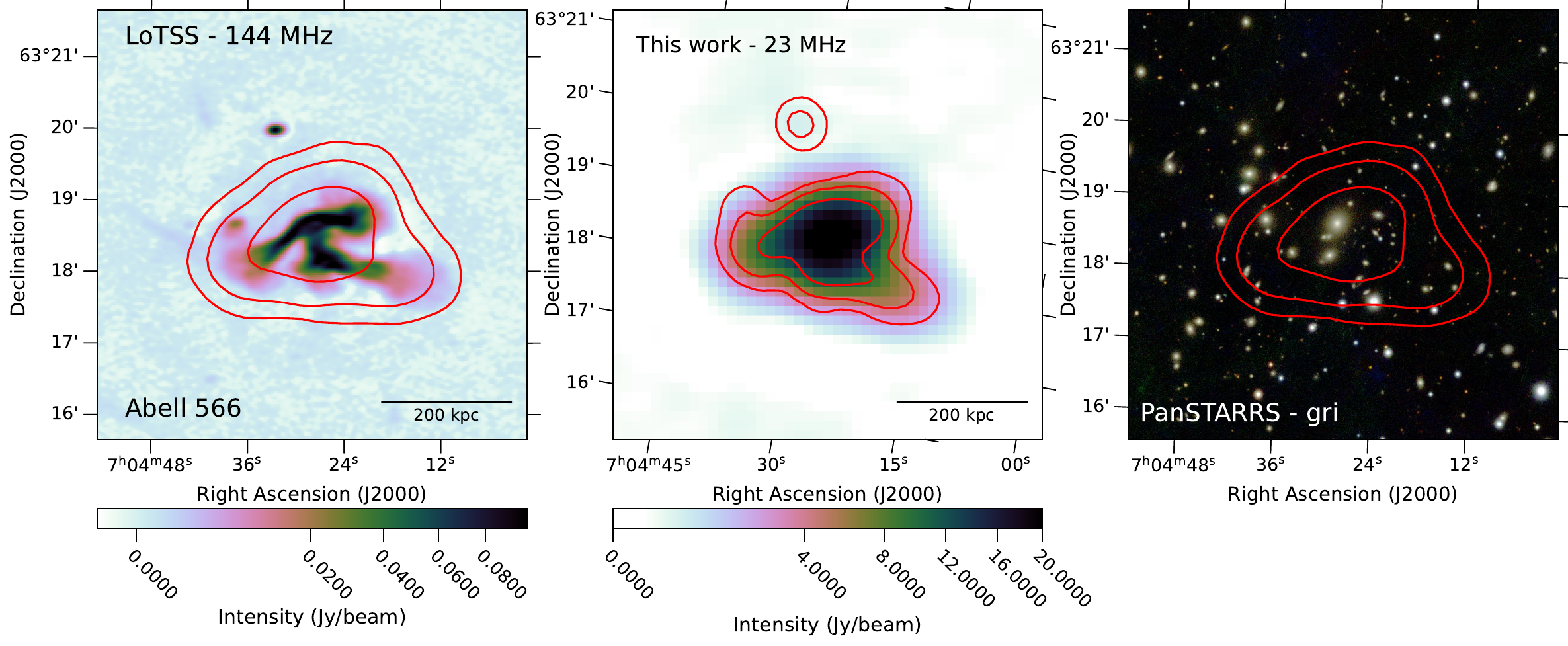}
    \caption{\small Images of four re-energised fossil plasma sources -- associated with Abell\,553 (\textit{top row}), LEDA\,2308099 (\textit{second row}), Abell\,565 (\textit{third row}) and Abell\,566 (\textit{bottom row}) -- detected at 23\,MHz. The left panel of each row contains a cutout made from LoTSS \citep{2022A&A...659A...1S} data, with 23 MHz contours overlaid at levels of $[8,16,32] \times \sigma$, with $\sigma=14$, 16, and 17 mJy~beam$^{-1}$ for the top, second and third row respectively (corresponding with the local noise). The center panel contains a cutout of the 23~MHz map presented in this work, with TGSS \citep{tgss} 151~MHz contours at $[1,2,3] \times \sigma$, with $\sigma =\SI{5}{\milli\jansky}$~beam$^{-1}$. The right panel contains a PanSTARRS \textit{gri} band \citep{panstarrs} / Legacy image, overlaid with the same contours as the left panel.
    Due to the brightness of the source, the bottom left image has 23 MHz contours overlaid at levels of $[32,128,512] \times \sigma$, with $\sigma=\SI{18}{\milli\jansky}$~beam$^{-1}$, and the bottom center image has TGSS 151~MHz contours at $[4,16,64] \times \sigma$.}
    \label{fig:cutoutsrc}
\end{figure*}

{To exploit the first sensitive decameter 23\,MHz source catalog, we searched for sources having extremely steep spectra. We did this by cross-matching our data with the 151\,MHz TGSS catalog and LoTSS. The TGSS survey covers the entire 305 square degrees area and achieves a similar depth to our 23\,MHz catalog for compact radio sources with a typical AGN spectral index of $-0.7$. %
The search reveals two extended sources, not detected in TGSS, but visible in LoTSS, see  \autoref{fig:cutoutsrc}. These sources also have the steepest spectral index among all the decameter sources detected in this work. 

The first source is located near the low-mass \citep[$M_{500}=\num{2.4e14}M_\odot$;][]{planck2} galaxy cluster Abell\,553 ($z=0.067$). Taking the 23\,MHz flux density and the 144\,MHz flux density from LoTSS, we compute an integrated spectral index of $\alpha_{23-144\,\rm{MHz}}=-1.82 \pm 0.22$. For this calculation (and the others ones reported below), the 144\,MHz flux density was extracted from a map tapered to the same resolution as the  23\,MHz map. The  steep-spectrum emission is located to the south and east of the brightest cluster galaxy (BCG). The source has a largest physical size of about 250\,kpc. The LoTSS image shows an irregular morphology, with the brightest emission to the south of the BCG. A separate compact radio source is associated with the BCG itself. 

The second extended steep-spectrum source is located near the galaxy LEDA\,2308099  \citep[$z=0.160$;][]{glade}. It has a spectral index of $\alpha_{23-144\,\rm{MHz}}=-1.43 \pm 0.22$. The LoTSS image reveals a complex horseshoe-shaped source, with the emission peaking to the west of LEDA\,2308099. The radio emission has an extent of about 430\,kpc. A faint compact radio source is associated with the galaxy. LEDA\,2308099 has no association with a known galaxy cluster or group. 

We also inspected resolved decameter sources with slightly less steep spectra. Most of these are classical radio galaxies. However, two of them have more peculiar morphologies. The first is located near the center of  Abell\,565, a low-mass cluster ($M_{500}=\num{1.6e14}M_\odot$) at $z=0.105$ \citep{mcxc}. 
For this 320\,kpc source, we measure a spectral index of $\alpha_{23-144\,\rm{MHz}}=-1.28 \pm 0.22$. The source was previously studied by \cite{MandalS2020Rtno}. Higher frequency flux density measurements show that the spectrum steepens significantly to $\alpha_{325-1400\,\rm{MHz}}=-2.05\pm 0.08$.  The second source is associated with the low-mass \citep[$M_{500} = \num{3.0e14}M_\odot$, $z=0.098$;][]{planck2} cluster Abell\,566. This complex 330\,kpc extended source has a steep spectrum with $\alpha_{23-144\,\rm{MHz}}=-1.26\pm 0.22$. It was also reported by \citep{MandalS2020Rtno} and seems to be associated with the BCG.

The irregular morphology, steep (curved) spectra, and presence of nearby radio AGN, but lack of clear direct association in the majority of cases, strongly suggest that the four sources discussed above trace revived AGN fossil plasma \citep{diffuse_gc}. 
Revived fossil plasma (also known in the literature as re-energised fossil plasma or radio phoenices) is predicted to be produced by the re-acceleration of plasma from older AGN outbursts. Simulations show that shock waves, for example from galaxy cluster mergers, can cause adiabatic compression of the old radio plasma \citep{fossil_plasma,2002MNRAS.331.1011E}. This produces radio sources with complex morphologies and steep, curved spectra. Very few such  sources have so far been identified  \citep[e.g.,][]{slee2021,mandal_fossil}, as their steep spectra make them faint at higher radio frequencies. Another expectation is that these sources are typically observed near cluster centers since the old radio plasma can be confined longer by the high density of the intracluster medium. Furthermore, fossil sources should also be common in low-mass clusters, in contrast to larger megaparsec scale diffuse cluster radio sources \citep{mandal_fossil}.
Both of these expectations agree with our finding that three of our four detected sources are located near the centers of low-mass clusters. The discovery of these sources, with two of them being the steepest 23\,MHz-detected source in the entire surveyed area, implies that fossil sources could form the dominant $\alpha\lesssim-1.5$ population in the decameter band at our sensitivity limit. It is thus expected that many more can be found when more area is surveyed. }
 \color{black}

From the results discussed in this work we conclude that sensitive sub-arcminute ground-based surveys of the decameter sky at frequencies above $\sim15$\,MHz are possible. Given the large field of view at these frequencies, such decametric surveys can be performed in a significantly shorter time than required for surveys at higher radio frequencies. Moreover, ground-based observations are an attractive option given that interferometers in space are costly. Finally,
by optimizing a telescope for direction-dependent ionospheric calibration, it might be possible to push the limits to even lower frequencies than what is currently achievable with LOFAR.
Previous work \citep{grotereber} has shown that in favorable conditions, the plasma cutoff frequency of the ionosphere can drop below 5~MHz.
Given that the observations discussed in this work were obtained under average ionospheric activity, such an optimized telescope could perhaps operate at frequencies as low as 5--10\,MHz during years around the solar minimum, when the ionospheric perturbations are the smallest.

\FloatBarrier

\subsection*{Methods}

\subsubsection*{Observation and initial data reduction}
For this work, an observation with LOFAR's Low Band Antenna (LBA) in the 16--30\,MHz frequency range was used.
The observation was performed after midnight, as the impact of Radio Frequency Interference (RFI) at decameter wavelengths is significantly more severe during daytime and early evening than after midnight (see \autoref{fig:flagged}).
We used a 5~hour dual-beam observation, with beams centered on RA:\,06h20m15s, Dec.:\,+66\textdegree 23m20s and on RA:\,06h17m19s, Dec.:\,+54\textdegree 20m49s; starting at 00:00 UTC, 11 November 2021.
Only the Dutch stations of LOFAR were used. %
With a third simultaneous beam, we observed the primary calibrator, 3C\,196, which is located 23\textdegree and 19\textdegree respectively from the center of the target beams.

The data were reduced in a multi-tiered approach. Firstly, we excised RFI from both our primary calibrator and target observation using AOFlagger \citep{aoflagger}.
Next, the signals from the bright off-axis sources Cygnus~A and Cassiopeia~A were removed from both the primary calibrator and the target observations \citep{demix}.
We then applied a pre-determined bandpass and polarization alignment solution to the primary calibrator, after which we performed a phase and amplitude calibration to remove the effects of the ionosphere and temporal gains. High-quality models of the primary calibrators sources at low frequencies are available for this purpose \citep{LBA_VLBI}. %
From the amplitude corrections, we determined the ``S4-index'' \citep{s4index}, which is a representation of the severity of the ionospheric scintillations.
The S4-index is defined as the standard deviation of the normalized signal intensity in the ionosphere $I(t)/\left<I\right>$, and higher values correspond with a more active ionosphere.
In order to verify that our observation represents a normal ionosphere, we analyzed observations of our primary calibrator on 41 other nights with different ionospheric conditions between October and December 2021.
The S4-index of these observations was between 0.09 and 0.5, with a median of 0.20 and a standard deviation of 0.10.
The S4-index of the observation presented in this work was $0.18$, which is comparable to the median value of the set of calibrator observations. 

Subsequently, we applied the same bandpass and polarization alignment solutions to the target fields, as well as the phase and amplitude solutions obtained from the primary calibrator 3C\,196.
This step effectively takes out the instrumental effects, leaving only ionospheric perturbations.
These perturbations have a clear frequency-dependent functional form.

Next, we self-calibrate the full field of view of each target pointing separately, using the 74\,MHz VLSSr \citep{vlssr} as a starting model, correcting for the average TEC effect across the full image and differential Faraday Rotation in the ionosphere.
Differential Faraday rotation is taken out by first transforming the visibilities to a circular basis, where this effect becomes a phase difference between the RR and LL correlations. We then solve for this phase difference using a point source model, employing \texttt{DP3} \citep{2018ascl.soft04003V}. This is followed by an antenna-based phase-only self-calibration cycle starting from the VLSSr model, constraining the phase corrections to be smooth along the frequency axis. This allowed us to make a preliminary image of the target field with WSClean \citep{wsclean_main}, facilitating the identification of the brightest objects in the field of view.

\begin{figure}[htb]
    \centering
    \includegraphics[width=0.95\linewidth]{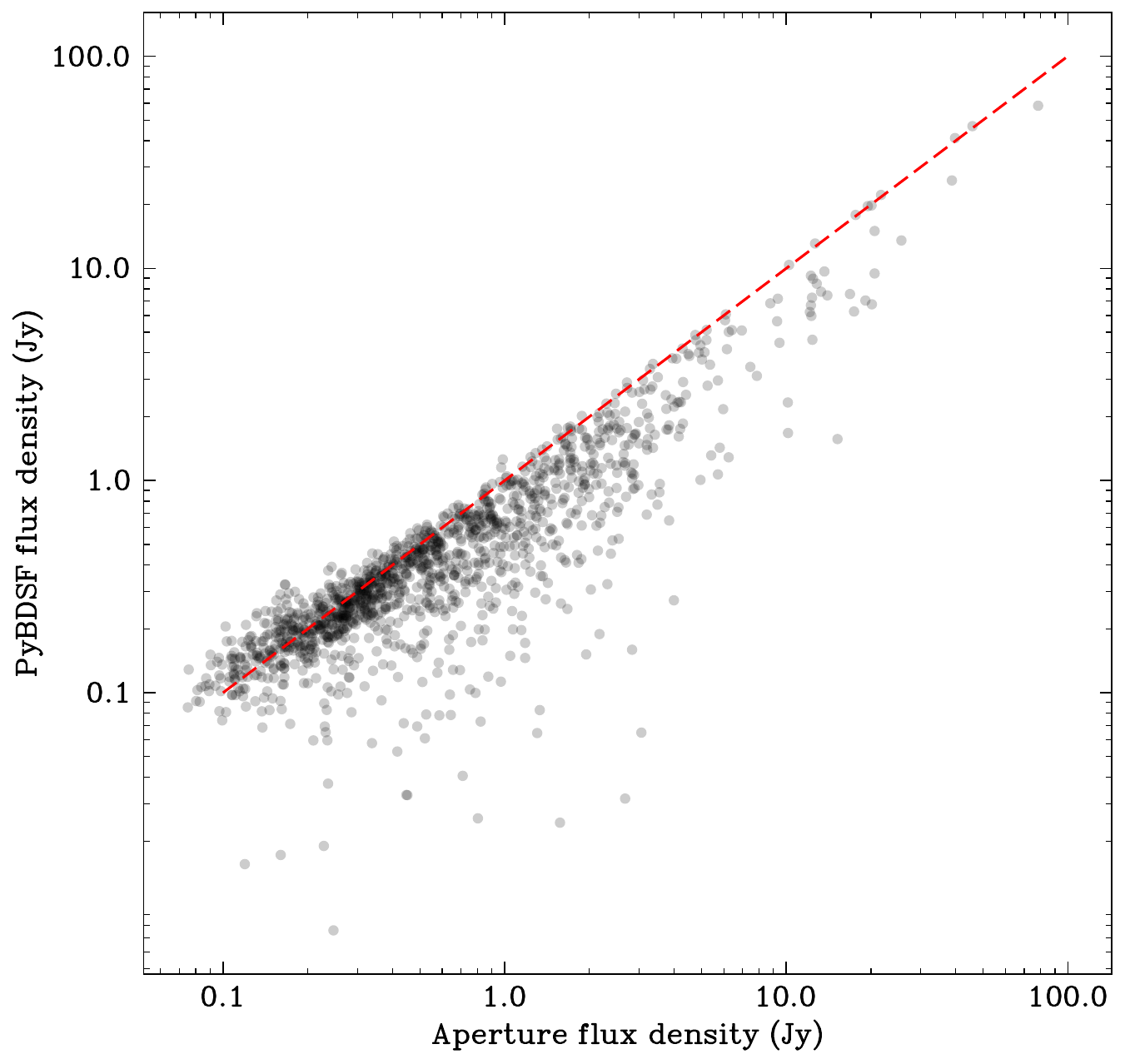}
    \caption{Comparison of flux densities measured with PyBDSF and with apertures (``aperture", these fluxes are reported in the attached catalog). The red dashed line represents the line where the PyBDSF flux densities are exactly equal to the catalog flux densities.}
    \label{fig:pybdsf_scatter}
\end{figure}

\begin{figure*}[htb]
    \centering
    \hspace{-0.05\linewidth}
    \includegraphics[width=0.5\linewidth]{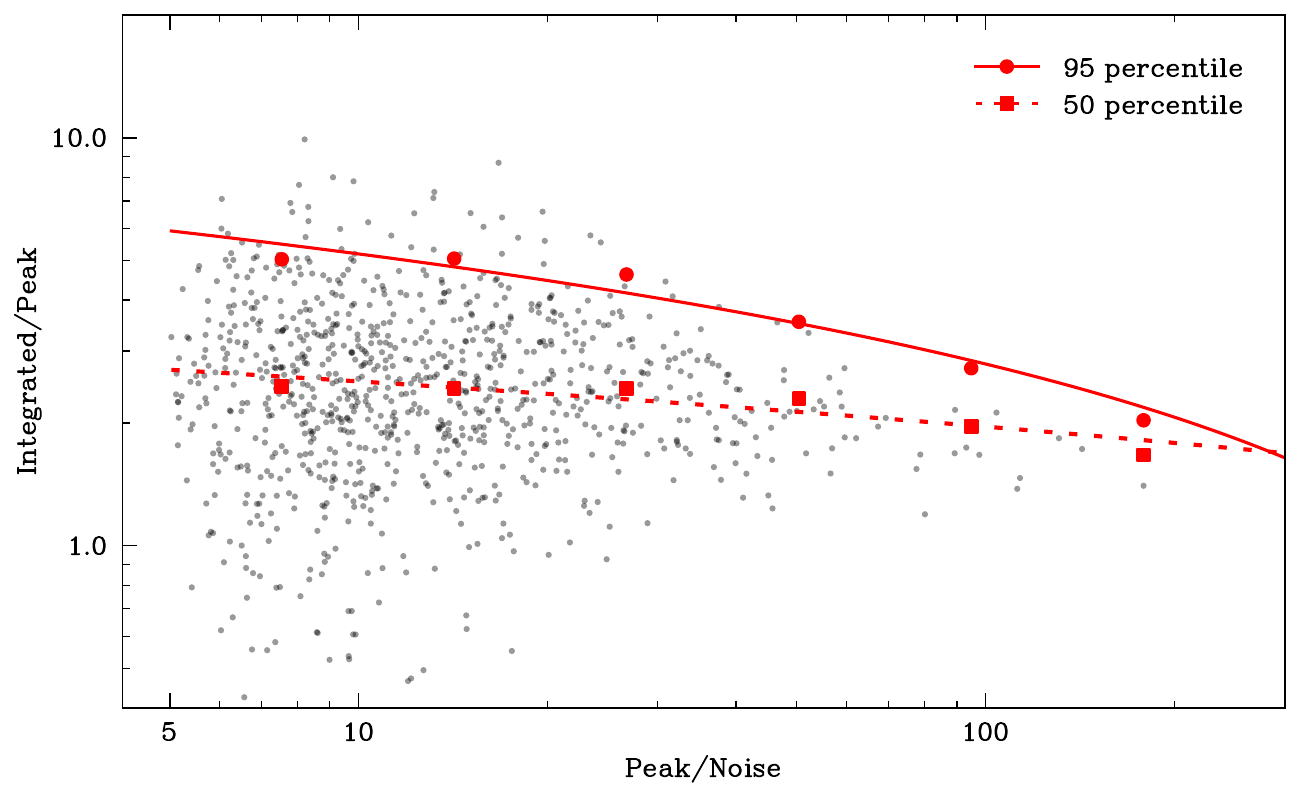}
    \includegraphics[width=0.5\linewidth]{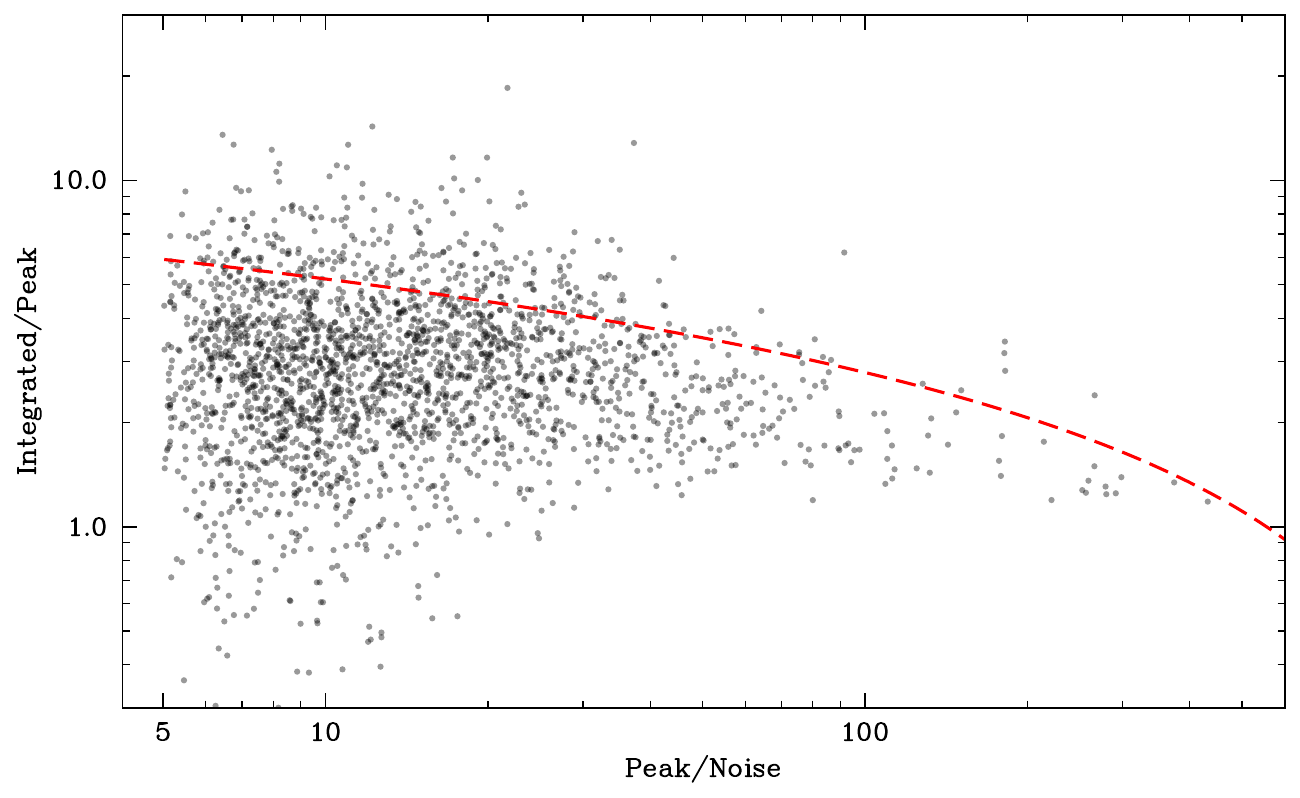}
    \caption{Integrated flux density per source divided by peak flux density, plotted over SNR of the peak flux density for all isolated sources (no other sources within 4 arminutes). On the left hand side, sources were additionally selected to be unresolved in the crossmatched LoTSS catalogue, which means that the blurring observed for these sources can only be explained with residual ionospheric effects. 
    The red points and squares indicate six bins where the 95 percentile and 50 percentile (=median) integrated over peak ratio is measured, with a line fitted to the six points. The 95 percentile line is also plotted on the right hand side - in this plot, 13.1\% of the sources lay above the 95 percentile line, indicating that an estimated additional 8\% of the sources are resolved in our map.}
    \label{fig:peakintegrated}
\end{figure*}

\subsubsection*{Direction-Dependent calibration}
As the ionosphere differs significantly across the field, most sources in the image with the direction-independent calibration are still significantly distorted. 
For this reason, the field is broken up in facets, which are subsequently self-calibrated,  with a technique similar to \cite{lolls} and \cite{facet_calibration}. For the self-calibration we make use of the \texttt{facetselfcal} script described in \cite{runwscleanlbauto}, solving for differential Faraday rotation and antenna-based phases with \texttt{DP3} that are constrained to be smooth along the frequency axis.
As a starting model, we use TGSS-ADR1 \citep{tgss} \footnote{This could also be LoLSS or LoTSS-based model, however, these were not yet available for the entire region of sky covered by our observation}.
Initially, this calibration step is only possible for facets containing the brightest few ($>10$\,Jy) sources, as fainter sources are severely perturbed by uncorrected ionospheric effects from the brightest nearby sources.
Using \texttt{DDFACET} \citep{ddfacet}, we are able to create an image and model, correcting each direction during imaging.
This improved model allows us to break up the field in more facets (about a factor 2 times more), as the ionospheric effects from bright sources are now better corrected.
The increased number of facets that can be calibrated results in an image that is sharp across a larger fraction of the field of view, allowing us to iteratively increase the number of facets that can be calibrated.
This cycle is repeated for in total 4 direction-dependent calibration cycles, giving us a total of 25 facets across the field of view of a single target pointing.

For each target pointing, the image was restored with a circular 45 arcsecond beam, which is close to the native resolution.
From this image, we extracted sources with \texttt{PyBDSF} \citep{pybdsf} into a preliminary catalog.
Due to residual ionospheric effects (particularly caused by residual scintillation and direction-dependent errors), unresolved sources are not well approximated by a Gaussian profile, which means that the fitted fluxes could be inaccurate.
For this reason, circular apertures centered on the positions obtained with \texttt{PyBDSF}, with a radius of 2 arcminutes are used to measure the flux for each source instead.
If sources are located within 4 arcminutes of each other, the corresponding apertures are truncated by the bisector of the line segment connecting the sources, ensuring that each source contains the flux corresponding to the sky area closest to the source.
Within each aperture, the total flux is computed by summing the intensity inside each aperture and dividing it by the area of the restoring beam.
Only sources with a ratio of peak flux density of 5 times the local RMS, an integrated flux density noise ratio above 7 and a maximum distance to the pointing center of 7 degrees were considered.
As this approach differs from the more conventional approach for creating radio catalogs \citep[e.g.][]{2022A&A...659A...1S,lollsDR1}, we compare the flux densities obtained in this work with the raw fluxes created with PyBDSF.
In addition, spurious sources located on calibration artifacts near bright sources were manually excised.
We see in \autoref{fig:pybdsf_scatter} that PyBDSF somewhat underestimates the flux densities of sources compared to aperture flux densities, however this underestimation is less pronounced for brighter sources, as the calibration solutions of the facets are weighted towards these brighter sources.
This is expected, as residual artifacts are not effectively fitted in PyBDSF but are included in apertures.

\begin{figure}
    \centering
    \includegraphics[width=\linewidth]{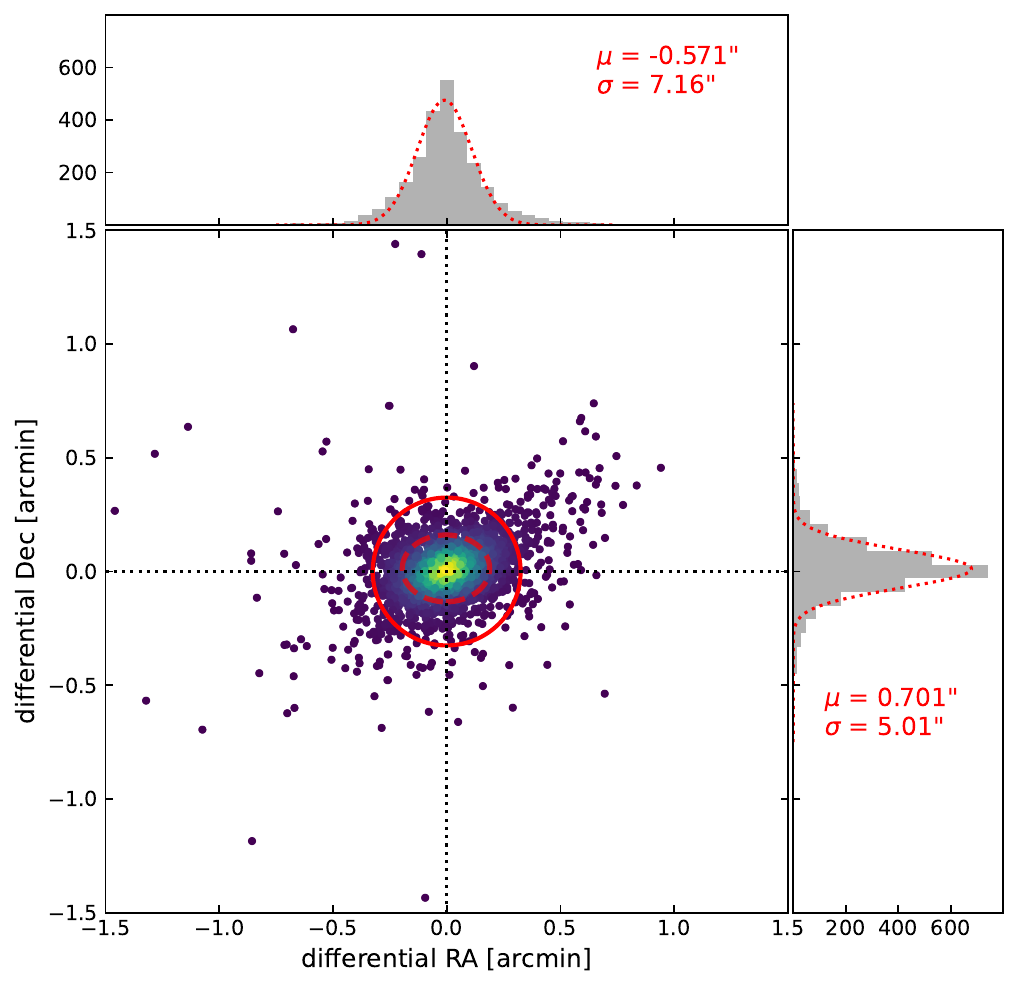}
    \caption{LOFAR source positions compared to TGSS, to verify astrometric accuracy. The color map represents the Gaussian kernel density estimation. The standard deviation of the offsets is indicated with the red dashed circle. The LOFAR beam is represented by the red, continuous line. This shows that the average astrometric offset is smaller than the restoring beam.}
    \label{fig:astrometry}
\end{figure}

The aforementioned residual ionospheric effects may cause blurring of sources, which causes a reduction in effective resolution. 
This results in a reduction of the peak flux density compared to the integrated flux density, as the blurring causes the full width at half maximum (FWHM) to be increased.
\autoref{fig:peakintegrated} shows the ratio of integrated flux density over peak flux density, where the ratio of integrated flux density over peak flux density corresponds to the broadening caused by residual ionospheric effects. For faint sources with a peak over local RMS ratio of around $\sim 5$, peak flux densities are reduced by a factor $\sim 2.45$ compared to the integrated flux densities, which means that the ionospheric blurring is around $\sqrt{(2.45\cdot 45)^2  - 45^2} = 100\arcsec$, while for bright sources, the ratio of peak flux density over local RMS is around $\sim 1.67$, which means that the ionospheric blurring is around $\sqrt{(1.67\cdot 45)^2 - 45^2} = 60\arcsec$.

\begin{figure*}
    \centering
    \includegraphics[width=\linewidth]{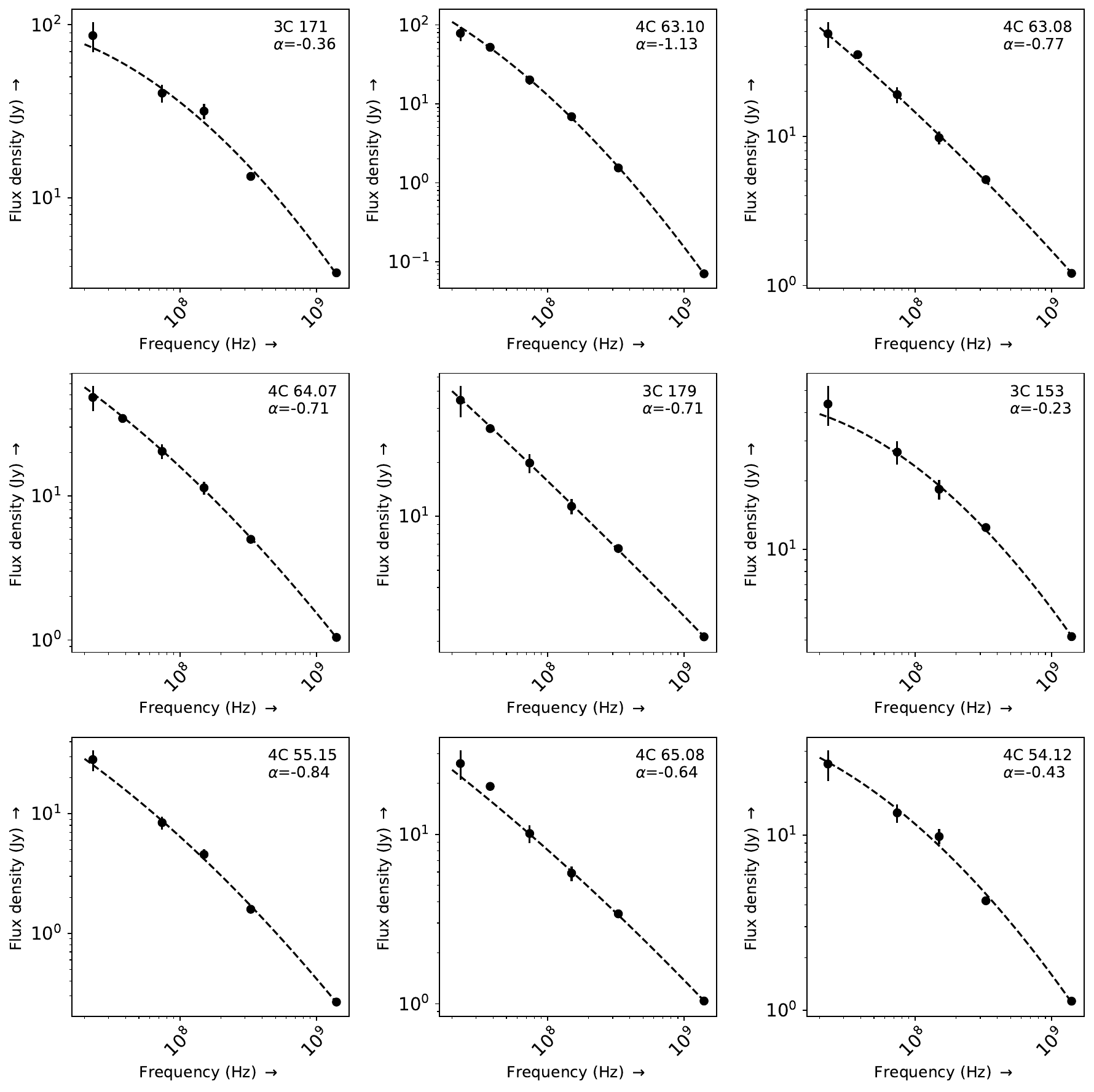}
    \caption{Spectra of the nine sources with counterparts in VLSSr \citep[74~MHz,][]{vlssr}, TGSS \citep[150~MHz,][]{tgss}, WENSS \citep[330~MHz,][]{wenss} and NVSS \citep[1.4~GHz,][]{nvss}, selected to have a flux density above 20~Jy in LOFAR. 8C \citep[38~MHz,][]{eightsee} fluxes are included if available, as 8C has only partial coverage of the sky area observed in this work. The data represents expected flux densities $\pm$ standard error. The dashed line represents a second-degree logarithmic polynomial, fitted through the data points of the 8C (if available), VLSSr, TGSS, WENSS, and NVSS surveys. Below the source name, the fitted spectral index at 23 MHz is reported.}
    \label{fig:comparison}
\end{figure*}

In order to verify the astrometric accuracy of our observation, we compared our catalog with the TGSS catalog at 151\,MHz, which has a comparable resolution to the data presented in this paper. 
In \autoref{fig:astrometry} it can be seen that the astrometric offsets are significantly smaller than the resolution of the LOFAR data, which indicates that the astrometry is reliable.
In \autoref{fig:comparison} we show the flux density of 9 sources with the highest flux density detected with LOFAR below 30~MHz, compared to previous source catalogs, to show that the flux density is in line with expected values given higher frequency data.
This agrees with \cite{lolls}, who also find that the LOFAR LBA flux-scale can be reliably transferred from the calibrator beam to the target beam.

\FloatBarrier

\subsection*{Acknowledgements}
CG and RJvW acknowledge support from the ERC Starting Grant ClusterWeb 804208.
MB is funded by the Deutsche Forschungsgemeinschaft (DFG, German Research Foundation) under Germany’s Excellence Strategy – EXC 2121 ``Quantum Universe" – 390833306.
EO  acknowledges support from the VIDI research programme with project number 639.042.729, which is financed by the Netherlands Organisation for Scientific Research (NWO).
AB acknowledges financial support from the European Union - Next Generation EU.
This publication is part of the project CORTEX (NWA.1160.18.316) of the research programme NWA-ORC which is (partly) financed by the Dutch Research Council (NWO).
LOFAR is the LOw Frequency ARray designed and constructed by ASTRON.
It has observing, data processing, and data storage facilities in several countries, which are owned by various parties (each with their own funding sources), and are collectively operated by the ILT foundation under a joint scientific policy.
The ILT resources have benefitted from the following recent major funding sources: CNRS-INSU, Observatoire de Paris and Université d’Orléans, France; BMBF, MIWF-NRW, MPG, Germany; Science Foundation Ireland (SFI), Department of Business, Enterprise and Innovation (DBEI), Ireland; NWO, The Netherlands; The Science and Technology Facilities Council, UK; Ministry of Science and Higher Education, Poland; Istituto Nazionale di Astrofisica (INAF), Italy.
The Pan-STARRS1 Surveys (PS1) and the PS1 public science archive have been made possible through contributions by the Institute for Astronomy, the University of Hawaii, the Pan-STARRS Project Office, the Max-Planck Society and its participating institutes, the Max Planck Institute for Astronomy, Heidelberg and the Max Planck Institute for Extraterrestrial Physics, Garching, The Johns Hopkins University, Durham University, the University of Edinburgh, the Queen's University Belfast, the Harvard-Smithsonian Center for Astrophysics, the Las Cumbres Observatory Global Telescope Network Incorporated, the National Central University of Taiwan, the Space Telescope Science Institute, the National Aeronautics and Space Administration under Grant No. NNX08AR22G issued through the Planetary Science Division of the NASA Science Mission Directorate, the National Science Foundation Grant No. AST-1238877, the University of Maryland, Eotvos Lorand University (ELTE), the Los Alamos National Laboratory, and the Gordon and Betty Moore Foundation.
When using data from the Legacy Surveys in papers, please use the following acknowledgment:
The Legacy Surveys consist of three individual and complementary projects: the Dark Energy Camera Legacy Survey (DECaLS; Proposal ID \#2014B-0404; PIs: David Schlegel and Arjun Dey), the Beijing-Arizona Sky Survey (BASS; NOAO Prop. ID \#2015A-0801; PIs: Zhou Xu and Xiaohui Fan), and the Mayall z-band Legacy Survey (MzLS; Prop. ID \#2016A-0453; PI: Arjun Dey). DECaLS, BASS and MzLS together include data obtained, respectively, at the Blanco telescope, Cerro Tololo Inter-American Observatory, NSF’s NOIRLab; the Bok telescope, Steward Observatory, University of Arizona; and the Mayall telescope, Kitt Peak National Observatory, NOIRLab. Pipeline processing and analyses of the data were supported by NOIRLab and the Lawrence Berkeley National Laboratory (LBNL). The Legacy Surveys project is honored to be permitted to conduct astronomical research on Iolkam Du’ag (Kitt Peak), a mountain with particular significance to the Tohono O’odham Nation.
NOIRLab is operated by the Association of Universities for Research in Astronomy (AURA) under a cooperative agreement with the National Science Foundation. LBNL is managed by the Regents of the University of California under contract to the U.S. Department of Energy.
This project used data obtained with the Dark Energy Camera (DECam), which was constructed by the Dark Energy Survey (DES) collaboration. Funding for the DES Projects has been provided by the U.S. Department of Energy, the U.S. National Science Foundation, the Ministry of Science and Education of Spain, the Science and Technology Facilities Council of the United Kingdom, the Higher Education Funding Council for England, the National Center for Supercomputing Applications at the University of Illinois at Urbana-Champaign, the Kavli Institute of Cosmological Physics at the University of Chicago, Center for Cosmology and Astro-Particle Physics at the Ohio State University, the Mitchell Institute for Fundamental Physics and Astronomy at Texas A\&M University, Financiadora de Estudos e Projetos, Fundacao Carlos Chagas Filho de Amparo, Financiadora de Estudos e Projetos, Fundacao Carlos Chagas Filho de Amparo a Pesquisa do Estado do Rio de Janeiro, Conselho Nacional de Desenvolvimento Cientifico e Tecnologico and the Ministerio da Ciencia, Tecnologia e Inovacao, the Deutsche Forschungsgemeinschaft and the Collaborating Institutions in the Dark Energy Survey. The Collaborating Institutions are Argonne National Laboratory, the University of California at Santa Cruz, the University of Cambridge, Centro de Investigaciones Energeticas, Medioambientales y Tecnologicas-Madrid, the University of Chicago, University College London, the DES-Brazil Consortium, the University of Edinburgh, the Eidgenossische Technische Hochschule (ETH) Zurich, Fermi National Accelerator Laboratory, the University of Illinois at Urbana-Champaign, the Institut de Ciencies de l’Espai (IEEC/CSIC), the Institut de Fisica d’Altes Energies, Lawrence Berkeley National Laboratory, the Ludwig Maximilians Universitat Munchen and the associated Excellence Cluster Universe, the University of Michigan, NSF’s NOIRLab, the University of Nottingham, the Ohio State University, the University of Pennsylvania, the University of Portsmouth, SLAC National Accelerator Laboratory, Stanford University, the University of Sussex, and Texas A\&M University.
BASS is a key project of the Telescope Access Program (TAP), which has been funded by the National Astronomical Observatories of China, the Chinese Academy of Sciences (the Strategic Priority Research Program “The Emergence of Cosmological Structures” Grant \# XDB09000000), and the Special Fund for Astronomy from the Ministry of Finance. The BASS is also supported by the External Cooperation Program of Chinese Academy of Sciences (Grant \# 114A11KYSB20160057), and Chinese National Natural Science Foundation (Grant \# 12120101003, \# 11433005).
The Legacy Survey team makes use of data products from the Near-Earth Object Wide-field Infrared Survey Explorer (NEOWISE), which is a project of the Jet Propulsion Laboratory/California Institute of Technology. NEOWISE is funded by the National Aeronautics and Space Administration.
The Legacy Surveys imaging of the DESI footprint is supported by the Director, Office of Science, Office of High Energy Physics of the U.S. Department of Energy under Contract No. DE-AC02-05CH1123, by the National Energy Research Scientific Computing Center, a DOE Office of Science User Facility under the same contract; and by the U.S. National Science Foundation, Division of Astronomical Sciences under Contract No. AST-0950945 to NOAO.
This work made use of EveryStamp\footnote{https://tikk3r.github.io/EveryStamp/}.
This work made use of Astropy:\footnote{http://www.astropy.org} a community-developed core Python package and an ecosystem of tools and resources for astronomy \citep{astro1,astro2,astro3}.

\subsection*{Author information}
\subsubsection*{Authors and affiliations}
\textbf{Leiden Observatory, Leiden University, Niels Bohrweg 2, 2333 CA Leiden, the Netherlands}\\
C. Groeneveld, R.~J. van Weeren, E. Osinga, J.~R. Callingham, T.~W. Shimwell, F. Sweijen, J.~M.~G.~H.~J. de Jong, G.~K. Miley, L.~F. Jansen, H.~J.~A. R\"{o}ttgering

\textbf{Netherlands Institute for Radio Astronomy (ASTRON), Dwingeloo, The Netherlands, Postbus 2, 7990 AA}\\
J.~R. Callingham, T. Shimwell, F. Sweijen

\textbf{Istituto di Radioastronomia, INAF, Bologna, Italy, Via Gobetti 101, 40129}\\
A. Botteon, F. de Gasperin, G. Brunetti

\textbf{Hamburger Sternwarte, University of Hamburg, Gojenbergsweg 112, D-21029, Hamburg, Germany}\\
F. de Gasperin, M. Br\"{u}ggen

\textbf{SKA Observatory, Jodrell Bank, Lower Withington, Macclesfield, SK11 9FT, United Kingdom}\\
W.~L. Williams

\subsubsection*{Contributions}
C.G. has coordinated and written the paper, reduced the data, and produced the LOFAR images. R.J.v.W. has developed the self-calibration strategy and led the proposal that provided the data this work is based on. E.O. has worked on the initial calibration strategy for decameter observations. F.d.G. has developed some of the procedures used for this work. J.R.C provided scientific background information on the physics of radio emission from stellar systems. W.L.W., J.R.C. H.J.A.R., M.B., G.B, G.K.M., and R.J.v.W. have helped with writing the paper and provided feedback on the manuscript. A.B. performed the LoTSS target extraction. T.S. has performed LoTSS data reduction and survey management. F.S. has had a critical role in developing the self-calibration software used in this work. J.M.G.H.J.d.J. has developed software required for managing direction-dependent correction files. L.F.J. has produced cross-matched catalogs between this work and LoTSS. W.L.W., J.R.C and L.F.J. have helped with verification of the data products.

\subsubsection*{Corresponding author}
Correspondence can be sent to \href{mailto:groeneveld@strw.leidenuniv.nl}{\UrlFont{groeneveld@strw.leidenuniv.nl}}

\end{document}